# Clustering of Indonesian and Western Gamelan Orchestras through Machine Learning of Performance Parameters

Simon Linke, Gerrit Wendt, Rolf Bader

**Abstract**

Indonesian and Western gamelan ensembles are investigated with respect to performance differences. Thereby, the often exotistic history of this music in the West might be reflected in contemporary tonal system, articulation, or large-scale form differences. Analyzing recordings of four Western and five Indonesian orchestras with respect to tonal systems and timbre features and using self-organizing Kohonen map (SOM) as a machine learning algorithm, a clear clustering between Indonesian and Western ensembles appears using certain psychoacoustic features. These point to a reduced articulation and large-scale form variability of Western ensembles compared to Indonesian ones. The SOM also clusters the ensembles with respect to their tonal systems, but no clusters between Indonesian and Western ensembles can be found in this respect. Therefore, a clear analogy between lower articulatory variability and large-scale form variation and a more exostistic, mediative and calm performance expectation and reception of gamelan in the West therefore appears.

## 1 History of gamelan transfer to the West

The early interest of Western scholars in gamelan and non-Western music, in general, was strongly related to the history of colonialism. Significant parts of Indonesia, including the Islands of Java and Bali, were colonized by the Dutch. Java was a colony for centuries; however, the courts of central Java enjoyed a certain amount of autonomy under colonial rule and became cultural centers where gamelan music thrived. (Sumarsam 1995) The neighbouring Island of Bali was colonised in different stages starting in the middle of the 19. Century. The whole Island came under Dutch control in 1908. The Dutch intervention led to the mass suicide of the Balinese royal houses of Badung and Klungkung. In contrast to Java, the Balinese courts played a minor rule as centres of the arts from then on.

The first known influence of Indonesian music and musical instruments in the West may have happened by the Hemony brothers, who made bells and compared the bell pitches with those of a metallophone made out of a series of metal rods (Kartomi & Mendonca, 2001, p. 505).This could be, but must not be, a kind of Indonesian metallophone. Also, Rameau owned a *gambang*, which he used for tonal studies. If these instruments were truly Indonesian is under debate, but quite likely.

So, although the flow of Indonesian art and culture to the West is much older, the first significant impact are the writings of Sir Stamford Raffles and John Crawfurd about the History of Java (Brinner, 1993; Kartomi, 1990). Being employed at the British colonial power in Indonesia (1811-1816), Raffles wrote a two-volume book about the archipelago, especially about the central Indonesian island of Java. He was also the first to bring a *gamelan* orchestra to London in 1830. Raffles and Crawfurd included depictions of the instruments and transcribed melodies using Western notation. Raffles bought two gamelan sets when he returned to Great Britain in 1816. He was also accompanied by Raden Rana Dipura, a Javanese nobleman, who performed on some of these instruments back in Britain (Bastin,

1971). This incident might be the first time a gamelan was played in Europe, even though not a whole ensemble but most likely only single instruments.

Raffles, who founded Singapore, was known to be a colonialist with a human face. I.e., he encouraged a form of farming preserving the natural resources and leaving part of the harvest to the native people to encourage them to continue their native living and so secure the colonial gain. This was 1830, a time of colonialism and growing interest in archaeology and peoples of the world. Still seeking European roots, it found its border in the Indus delta. Julius Braun, the founder of Comparative Archaeology, thought of the Christian occident as being founded by Greek culture, which he saw as a combination of the strict religious life of ancient Egypt and the more playful culture of Assyria, which was seen to be part of the culture of the Indus delta. (Braun 1856) This view is indeed much older and brought along by the philosophical and poetic tradition, mainly of Schlegel, but also of Hegel or Hölderlin in the times of German Idealism. So Indonesia was beyond that horizon and looked upon as a primitive culture in contrast to the high cultures of Greece and Egypt (and Europe, of course). So Raffles's writings about Java were not part of a movement of identification with ancestors and roots of the West but more part of the curious 'Bürgertum' interested in education. Here are the roots for modern Anthropology of that region and of what is called exotism, a projection of hopes, ideas, and feelings upon a culture, with interest in, but without a more profound knowledge about it and the necessity or willingness to go deeper.

The first substantial impact of Indonesian music on Western high culture music, what we call 'classical music' today, was the appearance of a *gamelan* orchestra at the Indonesian pavilion at the *Exposition Universelle* (world's fair) in Paris 1889. Here, and again in 1900, Claude Debussy heard a Javanese *gamelan.* However, it is unclear which type of gamelan (tuning, origin, style) was used on both occasions (Boyd, 1991). Taking part in the strong exotism present these days, he wrote 1905 his *Estampes* for piano with a piece entitled *Pagodes*. Here, he uses the equal-tempered pentatonic scales, which come close to the Javanese five-tone *slendro* scale. He also included other Chinese and Rag-Time influences in this Children's Corner Suite (1909). The use of *gamelan* melodies, rhythms, or timbres is not known in his work. This would probably be too far out of the taste as there was a clear perception of a border between music and noise.

A more important impact of Indonesian music on the West and also the starting point for the tradition of Western art composers writing music for or deeply influenced by *gamelan* was the work of Erich von Hornborstel and Curt Sachs, which founded the Deutsches Phonogrammarchiv in 1900. The first wax cylinder was a recording of a Thailand *phi-pha* orchestra (theatre ensemble from Bangkok, Siam) touring in Berlin these days and performing in the Berlin Zoological Gardens[1].

Hornborstel was educated as a composer and pianist, then studied natural sciences and philosophy, earned his doctorate in chemistry, and after moving from Vienna to Berlin, he worked on experimental psychology and musicology. He became Professor in Berlin in 1917 and was head of the Berlin Phonogramm Archiv till 1933 when he had to emigrate from Germany because of Nazi terror. Still, his foundations in musicology remained fundamental to the view of the discipline in the 20th century. They influenced the Western post-modern school of composers of art music. These include his work on tonal systems (Western and non-Western), music psychology and music perception, instrument classification with Curt Sachs in 1914, and maybe most prominently, his writings founding Vergleichende

---

[1] Various Artists (2000). *Music! The Berlin Phonogramm-Archiv (1900-2000)*.

Musikwissenschaft (Comparative Musicology) (Hornborstel, 1906). The idea of universals in music holding for all cultures was in contrast to pure diversity and meant to be a tool for understanding music and man alike.

Hornbostel tried to explain the origin of the tuning systems used in gamelan and other non-European musical traditions by a series of overblown fifths created on closed pipes. This idea became known as the theory of overblown fifth (Blasquitentheorie) and was supported by Jaap Kunst (Kunst, 1990), discussed below. So it was also in contrast to exotism and was searching for the fundamentals and not for 'the new, exotic'. (Hornbostel 1921)

Another vital researcher of ethnomusicology in Indonesia was Jaap Kunst. Educated as a violinist, he toured through the Dutch colony of the East Indies (now Indonesia) in 1919. Several cultural clubs of the ruling Dutch supported concert tours there. He had also studied law and was working in banking. Fascinated by Indonesia and especially the *gamelan*, he did not return then from his tour. He mainly stayed in Bandung and Batavia (now Djakarta) as a government employee in finance, leaving him time for his musicological studies. Until 1930, his work had attracted so much attention that he was given a full-time job as an ethnomusicologist. He was known to be highly active in the field, visiting small islands and remote areas and walking into small villages in the daytime when most people were out working in the fields. In Flores, he was climbing on a rock, starting to yodel (a vocal technique known on that island) (Cohen, 2016; Heins, 1978). People would come home from work and were keen on showing him their instruments, playing and singing. Hornborstel sent him new wax cylinders from Berlin while he sent his recordings back to Hornborstel. Rumor wants it that the first time he met Hornborstel at the Berlin central station, they agreed to recognize each other by playing Indonesian gongs loud and heavy. Kunst was forced to return to the Netherlands in 1934, became curator of the Royal Tropical Institute in Amsterdam, and began lecturing at the University in 1954.

Another influence of Indonesian music to the West was the film *Insel der Dämonen* (*Island of Demons*) by Victor Baron von Plessen, who had done fieldwork and films of and about Indonesia for 27 years, and Walter Spies in 1930/31 (released 1933). Spies was a gay German intellectual who, like many others, felt uncomfortable in Europe around the time of the First World War and was looking for an alternative lifestyle, new ideas, and also for other religious beliefs. The island of Bali seemed to fit these needs. One reason for that was the seemingly 'free' treatment of sexuality in Bali, where women typically did not dress the upper part of the body. This habit, which still was common for older women in Bali at the turn of the 21$^{st}$ century, was not meant to be related to sexuality by the Balinese. Still, for the Europeans, this was perceived as less restricted than the habits they knew from home[2].

Spies was raised in a wealthy German family in Moscow and knew many influential musicians and painters personally, like Rachmaninow or Skriabin. When moving to Dresden, he was introduced to Haba, Hindemith, Kokoschka, and Klee, next to others, who painted and played the piano professionally. He was also in the film business in Berlin, but in 1923, he fed up with it and decided to see something from the world. He went to Hamburg on a steamship called Hamburg with a friend, Heinrich Hauser, who was working as an ordinary seaman. Both were fascinated by the exotic[3].

---

[2]     Indeed, the use of naked breasts in paintings about Bali at that time was obligatory, in contrast to photographs of these times, where only a few women dressed like that (Hitchcock, M & Norris, L 1995) Here we have another form of exotism.

[3]     Hauer wrote to a Daisha, a sister of Spies from Colombo: 'Und dann nur Lichter und eine rießige Eingeborenenstadt mitten im Palmwald und Tempel mit Tänzen und eine phantastische unterirdische Musik, und

Spies deserted in Java from the ship work and went to Bandung to work as a pianist in a cinema and accompany singers and violinists touring through the East Indies (like Kunst has done). He then became conductor at the Kraton, the court of the sultan of Bandung, which was arranged by Maria Sitsen-Russer, head of the art circle 'Bond van Nederlandisch-Indische Kunstringen', who was fed up with the Javanese performance of the national anthem of the Netherlands at official receptions. Spies formed a band of Javanese musicians, teaching them violin and brass instruments, playing Jazz and Foxtrott (and the Netherland national anthem), which the sultan was very enthusiastic about. (Spies & Rhodius, 1964)

Here, Kunst started transcribing Javanese music for piano[4]. Unfortunately, he never published his transcriptions, although Hornborstel was urging him to do so. Spies later wrote Kunst that he refused to publish them as his transcriptions were a Western view on *gamelan* and so another thinking about that music, where several theories before had already failed and his would probably, too (Spies & Rhodius, 1964).

These activities were the starting point for transcriptions of this music for Western listeners, and later, Colin McPhee and Benjamin Britten played these transcriptions in several concerts in the USA. The first public appearance of Walter Spies and Colin McPhee playing four-handed piano was at the *Little Harmony Club* in Denpasar in 1937. Spies had settled in Ubud in Bali in 1927, which from then on was a world-known anchor point for wealthy and prominent travelers from around the world, i.e., he was visited by Charlie Chaplin, who stayed with him for several days.

As Ubud is in the south of Bali, and these days the entry point to Bali was the harbor of Singaraja (Bueleleng) in the very north, Ubud was a remote place then. Today, as the tourists come in at the airport near Kuta in the very south near Denpasar, the capital of Bali, Ubud can no longer be called quiet. Still, it is the center for Balinese art today, and Walter Spies lived there. The exotic of his perception of Bali was used in his film mentioned above. The poster advertising the movie in Europe shows a woman with bare breasts carrying a bowl of fruits on her head. In the background, a demon is shown. The subtitle is 'Zauber der Südsee. Bali. Tropische Wunderwelt. (Magic of South Pacific. Bali. Tropical Wonderworld.).

Musically, the film includes a scene of the *kecak* dance, a trance dance, where men are sitting in a circle, making a noise of repeating the syllables' keb - cac' at high speed and loud, and then again low in volume. Then, they fall into a trance, which is dramatically shown. Participants of that film scene, being very young then, reported at the end of the century that Spies did direct the scene and transformed the traditional *kecak*, which was much calmer and softer, into the more dramatic version for the sake of the film being more popular. As there are no recordings of the *kecak* before the intervention of Spies, we do not know how it was originally. Still, that dance and the music of *kecak* was for nearly a century associated with Bali much more than *gamelan* music and still today comes into people's minds.

---

die Luft ist dick von Brand und tausend Gerüchen und Kohlen in schwarzen Klumpen.' („And than just lights and a big city with natives in the middle of a palm forest and temples with dancers and a phantastic underground music and the air is thick of burnings and thousand smells and coal in black clusters.') (Spies & Rhodius, 1964)

[4] ‚Er konnte auf dem Klavier den *gamelan* so naturgetreu nachahmen, daß unsere Bedienten herkamen um zu lauschen, und entzückt sagten, daß es wie ein echter *gamelan* klänge.' („He could imitate the *gamelan* on the piano that naturally that our servants came along to listen and delighted said, that it sounded like a real *gamelan*.') (Spies & Rhodius, 1964).

Another strong impact of Indonesian music on the West was given by Colin McPhee, a Canadian composer and ethnomusicologist (Young 1986). He went to Bali from 1931-38 with an interruption in 1935-36 (McPhee, 1946). Like Spies, and supported by him, he settled in the Ubud area, enhancing the city to be the cultural center of Bali. McPhee wrote the first detailed book about the music and musical instruments used in Bali (McPhee, 1966) and, like Spies, wrote many transcriptions of *gamelan* music for piano, most of them found today at the McPhee archive at the Institute of Musicology at UCLA (Oja & McPhee, 1994). He, therefore, increased the understanding of *gamelan* music in the much more lively style of Bali compared to the Javanese *gamelan*, which was mainly perceived as 'the' Indonesian music.

## 2   Gamelan becoming a global musical tradition

During World War II, when Kunst was a curator at the Tropical Institute in Amsterdam, Babar Layar, the first gamelan group in Europe comprised of non-Indonesians, was formed by teenagers who heard gamelan music being played at the Institute. Kunst supported the group and supplied them with recordings and publications on gamelan. The group played through the 1950[th]. (Mendonca, 2011)

McPhee was part of an American movement of composers and ethnomusicologists interested in 'World Music', a term he knew from Henry Cowell and his lectures about the 'Music of the World' in New York. He became a professor of ethnomusicology at the University of California, Los Angeles (UCLA) in 1958, together with the University of Berklee, a center on that field, with California acting as a catalyst and melting pot for musics of the world, invited by the ethnomusicologist Mantle Hood.

The Musicologist Mantle Hood was a student of Kunst. In contrast to earlier researchers, Hood advocated for understanding and studying a foreign musical tradition through practically learning and performing it (Becker, 1983). At this point, it is essential to note that the development of Indonesian nationalism and, ultimately, Indonesian independence after World War II led to a democratization of the traditional arts (Cohen, 2016) This process made it possible for scholars such as Hood to go to Indonesia and practically learn gamelan compared to his predecessors. Hence, gamelan outside of Indonesia was no longer a pure exoticized musical tradition scientifically studied by Western scholars. It could be practically learned by foreigners abroad and within its culture sphere in Java and Bali.

Hood founded the first gamelan group in the United States at the University of California, Los Angeles. (UCLA). He was also the first to teach his students that music and everyone had to learn to play the instruments while studying Indonesian music from a theoretical and ethnomusicological standpoint. Hood introduced the term 'bi-musicality' to stress his view of a possible world music (Cottrell, 2007). This was in contrast to Kunst, who still was a colonial man who never would talk to Indonesian people directly or even play their musics himself.

McPhee wrote symphony music much influenced by *gamelan* music, where *Tabuh-Tabhuan* is his most widely known piece. He also performed his G*amelan* transcriptions with Benjamin Britten in several concerts in the USA. Britten, who also visited McPhee and Spies in Bali, knew *gamelan* music from shellac recordings before starting his trip (Mitchell, 1985). *The Prince of the* Pagodes is his most prominent piece influenced by *gamelan* music, using clearly the *saih selisir* (row of five), one of the most prominent Bali *gamelan* scales (Tenzer, 2000). Pentatonic scales were clearly associated with world music these days and 'the new exotic thing'. Britten also uses rhythmic patterns and instrumentation coming close to the timbre of the *gamelan* when the scene is entering pagoda land (Cooke, 1998).

The 'American Gamelan' is now primarily associated with Lou Harrison (Miller & Lieberman, 1999). He came from Seattle, Oregon, but settled in California and was influenced by the musical scene around the San Francisco Bay Area and the Musicology at UCLA. He was working on music of Mexico and the Indians there and was interested in other kinds of world music, listening to wax cylinders of Hornborstel and Kunst. He studied with McPhee at UCLA. He was then invited to Japan in 1961 for a conference about world music, followed by research in Japan and Korea. His knowledge of musical tunings came from his interest in old Greek music (Von Gunden, 1995). He was especially interested in just intonation, which he considered perfect and pure. He also applied this concept of simple ratios of musical intervals to the *gamelan* and built a set of instruments tuned in just intonation out of simple brass material.

This *gamelan* tradition continues and spreads throughout America, where more than 150 gamelan orchestras exist. The first gamelan built in North America is the 'Son of Lion' gamelan, founded by Barbara Benary, who constructed the instruments. Daniel Goode, a composer who had studied with Cowell and Philip Corner, wrote for the orchestra. Philip Corner, a representative of the Fluxus movement, composed about 500 pieces for gamelan, also using the Indonesian improvisation tradition (Lukoszevieze, 2003). Higgins describes his music in times and terms of Fluxus as the accumulation of events to a new Now and describes the intentions of Corners' music as a connection of new media (and musics) with an intention of meaning (Higgins Dick, 1969). Still, some kind of uncertainty remains as the principle to keep the music flowing. Despite these traditions, Dennis Murphy is called 'the father' of what today is called 'American gamelan'. Studying Ethnomusicology at the University of Wisconsin-Madison from 1959, he got in touch with Reed Tripp, then head of the economics department, who had brought along a small *gamelan* from Java. Murphy and Tripp started constructing their own set of instruments called 'Venerable Sir Voice of Thoom'[5] (Murphy, 1975).

In Europe, there are some historical gamelan orchestras (London 1830, Hamburg 1889). The tradition is most widespread in the Netherlands and England, with about 35 orchestras in each of these countries, but mostly everywhere orchestras can be found. In Germany, about ten active orchestras exist at the moment, Javanese and Balinese (Kebyar). One of the first orchestras in Germany is the Arum Sih orchestra of the Überseemuseum Bremen (Museum of Ethnology), founded in 1981 and much influenced by the New Age movement of the 80[th].
In the ensemble of the Indonesian Consulate in Hamburg native Indonesians play, dance, and sing with German musicians. The Consulate also engaged a professional Javanese musician who dances and performs the wayang kulit.

Additionally, the influence of *gamelan* led to other musical forms, most prominent the so-called Minimal Music of Steve Reich or Philip Glass. Also, in the music of King Crimson, clearly, a 'Minimal' input can be heard forming a 'Rock Gamelan'[6]. Unfortunately, a detailed discussion of these kinds of music is beyond the scope of this paper. Additionally, of course, there is also an influence of Western music in the region, from Portugesian music, Indonesian

---

[5] See also CD and paper of Judy Diamond about American *gamelan* or the webpage of the American Gamelan Institute http://www.gamelan.org/
[6] King Crimson IV: Discipline. The guitar sound of Robert Fripp reminds at that of Pat Metheny in his recording of a Minimal Music piece 'Different Trains' of Steve Reich.

Kroncong, Heavy Metal Bands in Bali[7], Skinhead Bands of Indonesia, or Nazi Bands in Singapore[8].

## 3 Differences between Indonesian and non-Indonesian Gamelans

Gamelan groups outside of Indonesia might, in some ways, differ from their Indonesian counterparts due to several factors. However, the differences are not as evident as they might seem at first glance. For members of those international Gamelan groups, their first attempt of playing the gamelan is often the first time they consciously listen to gamelan music. On the contrary, for players in Gamelan ensembles in Indonesia (Javanese, Sundanese, Balinese, etc.), it is much more likely that the players grew up to some extent by being aware and having listened to the music. However, even in Indonesia, Gamelan ensembles are very different depending on the origin of the players and the performance setting of the specific ensemble.[9] Besides that, the traditional knowledge of *karawitan* (Gamelan music) is not omnipresent in Java. In our experience, many Indonesian (including the Javanese) diaspora members have started to play the gamelan in Germany for the first time. On the other hand, some non-Indonesian Gamelan ensembles in Great Britain and the United States include many players who have studied at the Art Faculties in Indonesia (ISI). Naturally, their playing is quite advanced.

Gamelan beginners in the West usually start with the instruments playing the Skeleton-Melody *balungan,* such as the *saron*, *demung,* or *slenthem*. Many such players will not move on to the more challenging musical instruments, such as the *bonang*. In German groups, many saron and demung players exist, but there is a lack of players who can play the elaborating instruments, such as *gender*, *gambang*, *rebab,* etc. In many cases, this affects the choice of musical material since many categories of compositions have a more substantial presence of the elaborating instruments. Naturally, it also affects the ensemble's sound if specific instruments, such as the gambang or *rebab,* that provide a particular timbre are absent.

Different levels of musical dynamics characterize gamelan music. The balungan instruments especially use different loudness levels that might change during a performance according to the mood of a composition. The specific tempo is also related to the concept of *rasa*, a term not clearly defined in Javanese culture of Indian origin that can be translated to feeling or sense.[10] In Javanese gamelan, it relates to the mood and emotions a specific composition expresses. (Benamou, 2003). Musical dynamics are not included in the staff notation used to play gamelan music. Musicians must, therefore, know the musical piece and the musical culture in general to understand how to perform it, emphasizing musical dynamics and tempo.

---

[7] Most prominent and internationally recognized is the band *Superman is dead*, where ‚Superman' is meant to be Suharto, the former president and military leader of Indonesia.
[8] The spread of Nazi ideology in Indonesia is quite unknown. Just recently in a documentary film by James Leong & Jynn Lee about an appeasement meeting in Passabe, East Timor, where Indonesian military forced people of surrounding villages to kill people in Passabe, a band of keyboard and violin was playing. The keyboarder wore an old military jacket from the Indonesian army which had a swastika sticker on it, black swastika, white circular background with red outside, clearly a Nazi emblem. The director of the film, James Leong was also astonished about that and, asking the musicians and the people of Passabe, realized, that none of them knew about the meaning of the sticker (Leong, personal communication). Still the presence of this sticker on an Indonesian army jacket is alarming.
[9] Gamelans are often taught in high schools in Java or at Arts and neighbourhood centers (Sanggar). Locally those ensemble differ greatly in quality compared to ensemble comprised to some extent of alumni of the Art school (ISI, SMKI)
[10] The term has an Indian origin and is in the India also used to describe the athletics of different forms of art. However, according to (Benamou, 2003) the Javanese concept of rasa is much less defined than the Indian rasa-theory.

The ability to play a musical piece in a specific tempo and loudness related to the suitable *rasa* is another aspect that affects the sound and might differentiate Indonesian and non-Indonesian Gamelans.

## 4 Gamelan Tonal Systems

The discussion of *gamelan* music in the West also elaborated on pitch and timbre. Pitch was found interesting in two ways. First, compared to Western music, the pentatonic scale sounded archaic and exotic. Secondly, the temperament was new. The scale Benjamin Britten used in *The Prince of the Pagodes* is *not* the Balinese scale but one being put into the shape of Western equal tuning. Then it reads like a major scale with the second and sixth steps missing. When played to Western listeners familiar with Balinese music, the association with Balinese music is clearly there if the melody is rhythmically simple, too. The tuning of *gamelan* differs from village to village. Spies had the idea that it was a compromise between different instrument tunings and needs, which he mentioned to Hornborstel in a letter. Still, we do not know precisely about his ideas.

Only recently has a system of older and more modern tuning shown up for Balinese tuning (Tenzer, 2000). Here, the old tuning called *tirus* consists of steps (0, 197, 377, 724, 828, 1200 cent), which would come close to a major scale with no fourth and seventh but with a minor sixth. The more modern scale *begbeg* scale (0, 120, 234, 666, 747, 1200 cent) shows much more narrow intervals, which could be heard as minor and major second, sharpened tritone, and sharpened fifth. Still, there is a middle tuning called *sedeng* (0, 136, 291, 670, 804, 1200 cent), where the intervals are not as sharp as with *begbeg* and not as big as with *tirus*. It could be a minor scale with a flat fifth and a sharp second. In all scales, there is no fourth and a seventh.

The discussion about the tuning of *gamelan* instruments in Western literature is quite large. Psychologically, this is a matter of pitch perception as categorical perception. Western tonal music knows twelve discrete pitches in one octave. If tones that do not fit the precise tuning of Western scales are presented, listeners try to use this pattern as a schemata. Tones that do not exactly meet the expected points in the tonal space are still heard as these tones but are perceived as slightly detuned. This makes it possible for Western listeners to fit the *Gamelan scales into Western scales and* play Western instruments with *genders* or *sarons*[11](Schneider, 1997a).

## 5 Gamelan Timbre

The *gamelan* timbre differs from timbres of western instruments. The metallophone rods do not show a harmonic overtone series as i.e., string or blown instruments have. Western listeners are used to sounds with harmonic series. Psychoacoustically, harmonic sounds are treated by the human brain in a way of fusion (Schneider, 1997b). The different partials of the sound cannot be heard separately and distinguish one from another. They rather fuse to only one tone sensation. The amount and amplitude of the partials of a harmonic sound do only contribute to the overall 'tone color' of the fused sound.

---

[11] One example of this are the works of Lou Harrison for gamelan and orchestra i.e. the Double Concerto for Violin and Cello with Janavese Gamelan (1981-82) or Gutama Soegijo and the banjar gamelan group of Berlin playing with saxophonist Detlef Bensmann.

On the other hand, percussion instruments do not have such a harmonic overtone series. The reasons for this are manifold and far beyond the scope of this paper (Rossing, 2000). Still, western percussion instruments try to get as close as possible to such a harmonic series. This normally means the tuning of between one to about five partials in a series of simple numerical relations. Examples are church bells, where two octaves appear with a minor third (Rossing, 2000). Also, the xylophone bar is tuned so that the second partial is an octave above the fundamental. Some xylophones even tune a third partial to a major or minor third (or in between) above this octave.(Borg 1983) All these tunings are achieved by changes in the geometry of the instruments. So, the shape of the bell is the way it is to achieve this very set of overtones. The xylophone bar has its cut-off not only to get smaller bars for lower pitches, but exactly this cut-off tunes the partials. So, even with percussion instruments, western instrument builders try to achieve as much harmonicity as possible. Still, the higher partials are inharmonic and produce a sound always associated with percussion instruments.

*Gamelan* metallophones, on the other hand, do not try to obtain a harmonic overtone series. This is astonishing at first because *gamelan* builders care much about the metal bars' precise sound, pitches, and musical beatings. I.e., the large gong has a so-called *ombak*, an amplitude modulation of the fundamental partial, a beating that is between two and four beats per second (Schneider, 1997b). This also holds for the *gender wayang*, where two pairs of *gender dasa* (ten-plate metallophone) play together. Each pair is tuned so that the pitches differ again between two and four Hz, and so cause a beating with that frequency difference. But tuning the gender plates to produce a harmonic overtone series is not intended and not found with Javanese and Balinese metallophones.

Still, one perceives a clear pitch when the bars play a melody. This pitch is the bar's fundamental pitch, which is boosted by the bamboo resonators. Nevertheless, one can also hear additional pitches of the lowest partials and may sing the melody with these pitches. This would not be possible with Western percussion instruments as these lowest pitches are harmonically tuned, and so fuse with the fundamental partial, and the higher partials then are that high and damped out too fast that one could still perceive a percussion sound but is hardly able to hear a clear second or even third pitch, which is possible with *gamelan* metallophones. So, these bar instruments do have a certain amount of ambiguity. A beautiful listening example is the recording of Colin McPhee's gamelan transcriptions for four-hand piano, which he plays with Benjamin Britten.(Various Artists, 1999, tracks [19-25]) Although there are other publications of these piano pieces, the mentioned CD contains the original pieces played by a *gamelan* orchestra. One hears the difference, where the *gamelan* sounds more ambiguous, and the piano pieces are listened to more definitely and strictly[12].

But the *gamelan* sound also differs tremendously between Java and Bali, as do the gamelan styles. The more Islamic island of Java, with a *gamelan* also connected to a court tradition, has a much softer style than in Bali. Still, the Balinese *gong kebyar* is an invention of the 20$^{th}$ century in the region of Bueleleng in the north of the island under the impression of the Dutch conquering the last free kingdoms of Bali in 1906 and 1908(Tenzer, 2000). Here, the royalties, when recognizing their desperate and hopeless situation, walked against the Dutch only to commit suicide before the eyes of the soldiers. The shock of this event led to a search for the identity of the Balinese, which was then found in old writings of the Hindu *Ramayana* and *Mahabharata* epos. Reading competitions accompanied by the *gamelan* led to the invention of the *gong kebyar*, which is much faster and more complex than i.e., the older *gong g'dé*. This could be interpreted in terms of a Western impact leading to this new style. The Javanese gamelan metallophones are played mainly by soft mallets, producing nearly no overtone series and almost no inharmonicity. The saron, played by a wooden hammer, has plates so thick that the damping of the higher harmonics is so strong that they are nearly nonexistent. Still, some inharmonicity is left, which one can clearly perceive.

---

[12] The piano on this recording is also tuned in an equal tempered western manner, still the sound difference is obvious. Spies had tuned a piano in Bali in a way, that the white keys form a heptatonic and the black keys a pentatonic scale. Then he was able to play saih selisir and saih pitu originally. At the performance in the *Harmony club* in Denpasar with Colin McPhee an official Balinese stood up after the performance, enthusiastic about the transcriptions of the melodies but sorry, that the tuning was wrong. This piano was still tuned in a western manner.

On the other hand, the Balinese metallophones are normally played with a wooden hammer and hard attack, producing a wide range of overtones. The precision of the craftsmanship of Balinese *gamelan* builders can also be observed in *the gender wayang dasa* plates. Here, the trapezoid shape enhances additional inharmonic partials, which are much more than they would be with a flat plate (Bader, 2004)[13]. So here, the care in instrument building also holds for sound – not just for pitch - and there are no attempts to form a harmonic overtone series. This can also be associated with the different cultural world views(Bader, 2002). While the Western tradition has the idea of simple numerical ratios governing music and the cosmos in the philosophy of Platon[14] or Pythagoras, Hindu culture is not that hierarchical, and in their pantheon, many gods are present everywhere in nature all the time, leading to a huge diversity and ambiguity. This ambiguity is also present within the music, and as it does not conflict with the world view of Balinese Hinduism, there is no need to form a harmonic overtone series. This is slightly different in Java. Although Islam has rather critical views of music, in Indonesia, different types of music are performed during Islamic religious feasts. So, the strictness of this prohibition is not as strong as in Arabian cultures. Still, it seems to have influenced the music and may be the reason for the much softer sound of the Javanese *gamelan*.

Now, in the Western musical tradition of the 20th century, sound and timbre were increasingly getting into the focus of composers and listeners. This was taken even further by the new technical possibilities used in popular music, techno, or electronic music. Also, in art music, i.e., with Ligeti's famous *Atmospheres*, timbre became the focus of composition. The Indonesian music, therefore, was not only a new pitch system but also a new sound. It meant the dissolution of traditional harmonicity not only in tonal theory but also within the sound itself. The diversity of these sounds, their beating character, and their ambiguity of several perceivable pitches within one played note fit very well in the search of Western music for new timbre possibilities in the 20th century. Therefore, metallophone ensembles from Southeast Asia are unique in that although the sounds are diverse, the instruments are still carefully crafted and do not sound randomly but rather ambiguous.

## 6 The ensembles used in this study

The 58 recordings of non-Indonesian Gamelans used in this study are mainly comprised of Gamelan groups from Germany. They are listed in Table 1. The Indonesian recordings were taken from several Compact Discs featuring some of the well-known Gamelan ensembles from Indonesia. By using recordings of real gamelan ensembles, the present study stands out in comparison to previous approaches on the topic, as the perception of Gamelan music has been investigated with different results when using single synthetic sounds (Marjieh et al., 2024) in comparison to real instrument sounds in a musical context (Wendt & Bader, 2019).

---

[13]    Simulated sounds for flat and trapezoid plates can be listened to at https://rolfbader.de/physical-modeling.
[14]    Most prominently in his dialog ‚Timaeus', the so called ‚Thimaeus scale'.

**Table 1** List of the musical pieces and the performing ensembles of the recordings used for further analysis

| Title of piece | gamelan ensemble |
|---|---|
| Swa Buana Paksa | Gamelan STSI Bali |
| Tabuh_Pat_Jagul | Gamelan STSI Bali |
| Gabor | Gamelan STSI Bali |
| Kebyar_Duduk | Gamelan STSI Bali |
| Cita Utsawa | Gamelan STSI Bali |
| Topeng_Arsa_Wijaya | Gamelan STSI Bali |
| Teruna Jaya | Gamelan STSI Bali |
| Gendhing Gambir Sawit | Pendopo Gamelan ISI Surakarta |
| Gendhing Bonang Dhenggung Turulare | Pendopo Gamelan ISI Surakarta |
| Gendhing Mandulpati | Pendopo Gamelan ISI Surakarta |
| Gendhing Carabalen | Pendopo Gamelan ISI Surakarta |
| Gendhing Kemanak Bedhayan Pangkur | Pendopo Gamelan ISI Surakarta |
| Ketawang Puspawarna | Gamelan Paku Alam |
| Gending Tejanata / Ladrang Sembawa | Gamelan Paku Alam |
| Gending_Mandulpati_Ladrang_Agun-Agun | Gamelan Paku Alam |
| Bubaran_Hudan_Mas | Gamelan Paku Alam |
| Ladrang Turun Sih | Kyai Kaduk Manis Manis Rennger |
| Sekaten Gendhing Rambu | Kyai Kaduk Manis Manis Rennger |
| Ladrang Mijil Ludira | Kyai Kaduk Manis Manis Rennger |
| Sekaten Gendhing Rangkung | Kyai Kaduk Manis Manis Rennger |
| Ketawang Mijil Dhempel | Kyai Kaduk Manis Manis Rennger |
| Ketawang Puspawarna | Gamelan Mangkunegaran |
| Gending Bonang Babar Layar | Gamelan Mangkunegaran |
| Gending Ela-Ela Kalibeber | Gamelan Mangkunegaran |
| Ayak-Ayakan Kaloran | Gamelan Mangkunegaran |
| Unknown piece | Sekar Kenanga |
| Manggung Sore, Ricik-ricik | Sekar Kenanga |
| Mugi Rahayu | Sekar Kenanga |
| Unknown piece | Sekar Kenanga |
| Unknown piece | Sekar Kenanga |
| Lancaran Gendrowo Momeng | Arum Sih |
| Ladrang Didradameta | Arum Sih |
| Ladrang Wilujeng | Arum Sih |
| Agen-Agen, Gangsaran, Srepegan | Arum Sih |
| Kodhok Ngorek | Arum Sih |
| Perang Kembang | Arum Sih |
| Ladrang Pangkur | Arum Sih |
| Lancaran Tropongbang, Ladrang Bedrug | Arum Sih |
| Lagu Swara Suling | Arum Sih |
| Ladrang Wilujeng | Arum Sih |
| Ladrang Jatikumara | Arum Sih |
| Ladrang Surengrana | Arum Sih |
| Lancaran Serayu | Arum Sih |
| Ladrang Gonjang-Ganjing | Arum Sih |
| Tari Prawira Watang (Prajuritan) | Arum Sih |
| Langgam Kadhung Tresna | Arum Sih |
| Bubaran Udan Mas | Arum Sih |
| Ketawang Subakastawa | Margi Bodoyo |
| Ladrang Wahyu - Lancaran Mikat Manuk | Margi Bodoyo |
| Langgam Caping Gunung | Margi Bodoyo |
| Margi Bodoyo | Margi Bodoyo |
| Ketawang Manggung Sore - Ricik Ricik Banyumasan | Margi Bodoyo |
| From Shadows | Taniwha Jaya (BaliNewZealand) |
| Delirious Euphoria | Taniwha Jaya (BaliNewZealand) |
| Headrush | Taniwha Jaya (BaliNewZealand) |
| SZUNN | Taniwha Jaya (BaliNewZealand) |
| Padhasapa | Taniwha Jaya (BaliNewZealand) |
| Tinggal | Taniwha Jaya (BaliNewZealand) |

## 6.1 Non-Indonesian Ensembles

Four different non-Indonesian Ensembles are considered in this study. They are mainly German and listed in Table 2.

**Table 2** List of analyzed non-Indonesian gamelan ensembles and their origin

| Non-Indonesian Gamelan ensembles | origin |
|---|---|
| Margi Bodoyo | Hamburg (Germany) |
| Sekar Kenanga | Hamburg (Germany) |
| Arum Sih | Bremen (Germany) |
| Taniwha Jaya (BaliNewZealand) | Wellington (New Zealand) |

Arum Sih was most likely the first gamelan group to be established in Germany. The group was founded in 1981 at the Übersee-Museum in Bremen (Museum of Ethnology, Trade, and Natural History) under the guidance of ethnomusicologist Andreas Lüderwaldt (Lüderwaldt, 2007).

The group received introductory lessons from the Dutch ethnomusicologist Ernst Heins, who also assisted in ordering the musical instruments in a gamelan workshop in Surakarta (Solo). Besides introductory workshops, the group members mainly taught themselves over the years and went to Indonesia together in the 1990s. The ensemble consists of a complete double set of instruments, one in Slendro and one in Pelog. The group also played modern compositions by Western composers such as Bill Alves and Lou Harrison. The recordings of *Arum Sih* used in the present study were taken from two Compact Discs recorded in the 1990s and early

2000s. They are comprised of traditional Javanese music, mainly short categories of musical compositions such as *Lancarans*, *Bubarans,* and *Ladrangs*.[15]

Margi Bodoyo is a gamelan group from Hamburg. It was established at the Indonesian Consulate (KJRI) in Hamburg in the early 2000s and is led by Pak Maharsi, a Dancer, Dalang[16] , and gamelan musician from Semarang. The group performs regularly at events related to the Consulate. Its players are partly comprised of non-Indonesians and members of the Indonesian Diaspora in Hamburg. The history of the Instruments of Margi Bodoyo is somehow obscure. The Consulate owned a slendro set of Instruments before the establishment of the group. The pelog set is younger. Its origin is unknown. Over the years, Pak Maharsi has returned some of the Instruments and modified single saron-like Instruments that belonged to the Consulate, such as a *saron* from a Betawi-style (Jakarta) gamelan.

The group plays classical Javanese gamelan music, often with a focus on the cassette era of the 20th century. Pak Maharsi taught the group to accompany a short one-hour version of the shadow puppet theatre Wayang Kulit (often in German), with him as a Dalang. For that purpose, he has simplified the accompaniment with a standardized notation. The audio files by Margi Bodoyo are comprised of field recordings made during performances ranging from 2014 – 2020.

Sekar Kenanga is a gamelan group from Hamburg that was founded in 2017 by the participatory program of the Elbphilharmonie, Hamburg's concert hall close to the Elbe River. The group is led by Steven Tanoto, a gamelan musician from Hamburg with Indonesian roots, and is mainly comprised of non-Indonesians. The Sekar Kenanga Gamelan was made in the 1940s for a rich Javanese landowner. In 1999, it was bought by a French collective and stored in a village in southern France before being bought by the Elbphilharmonie in 2017 (Schulz, 2017). Sekar Kenanga plays traditional Javanese music and performs at least twice a year at the Elbphilharmonie in Hamburg. The concerning audio files were taken from field recordings made during rehearsals as well as performances.

Gamelan Taniwha Jaya is a Balinese ensemble from the New Zealand School of Music in Wellington. The specific recordings are part of a Compact Disc that features non-traditional compositions for Balinese and Javanese Gamelan.

### 6.2 Indonesian Ensembles

Five different Indonesian Ensembles are considered in this study and listed in Table 3. The recordings were taken from several commercial Compact Discs featuring some of the most representative gamelan ensembles. Those include two recordings, Javanese Court Gamelan Vol. I and II from the Nonsuch Explorer series were originally published by Nonsuch. Vol. It was recorded by ethnomusicologist Robert E. Brown and is one of the most famous, if not the most famous, records of gamelan music ever produced. It features music from the Paku Alam palace in Yogyakarta. The ensemble heard on the recording is led by K.R.T Wasitodiningrat, one of the most famous gamelan musicians of the 20[th] century. (Heins, 1978) Its opening piece, *Ketawang Puspawarna,* is featured on the Golden Record and included in the Voyager Spacecraft launched in 1977. Vol. II focuses on the music of the Mangkunegaran palace in

---

[15] Arum Sih 2001  
[16] The Puppeteer of the shadow puppet Theater Wayang Kulit

Surakarata. It features the Gamelan Kai Kanjut Mesem, one of the most famous gamelans from Central Java.[17]

**Table 3** List of analyzed Indonesian gamelan ensembles and their origin

| Indonesian Gamelan ensembles | origin |
|---|---|
| Kyai Kaduk Manis Manis Rennger | Surakarta (Java) |
| Gamelan Paku Alaman | Yogyakarta (Java) |
| Gamelan Mangkunegaran | Surakarta (Java) |
| Pendopo Gamelan ISI Surakarta | Surakarta (Java) |
| Gamelan STSI Bali | Denpasar (Bali) |

To other recordings by Yantra Productions Gamelan of Central Java Vol IV. (2004) and Vol VIII. (2007) where included. Both feature musicians from the Art school STSI and the Palace in Surakarta. They were recorded in the early 2000s and produced by John Noise Manis. Both Compact Discs mainly feature traditional Javanese Gendhing in Surakarta Style. Vol. IV. focuses on music related to Ritual and Spiritualism, such as a musical piece of the Gamelan Sekaten, a gamelan played once a year during the Islamic Sekaten Festival. Gamelan of Central Java vol. VIII. was recorded in Surakarta's palace (Kraton) and features the Gamelan *Kyai Kaduk Manis Manis Rennger*. Vol IV. was recorded on the Pendapa Gamelan of STSI Surakarta (The School of the Arts in Solo). Music of the Gamelan Gong Kebyar vol. 1 (1996) was included as a representation of Balinese Gamelan. The recording features Gamlen Gong Kebyar's music, an ensemble type developed in the early 20$^{th}$ century but now dominates Balinese gamelan music. Its music is characterized by sudden changes in tempo and dynamics and its bright, shimmering timbre. (Tenzer, 2000) The ensemble is led by *Wayan Berata*, one of the genre's most famous teachers
and musicians.

## 7 Sound Analysis Methods

The machine learning (ML) approach of self-organizing Kohonen maps (SOM) is used to analyze the musical pieces. Self-organizing Kohonen maps have been proposed for pitch and chord mapping (Leman, 1995; Leman & Carreras, 1997), or sound level assessment (Kostek, 2005); for a review, see (Bader, 2013). Regarding Computational Ethnomusicology, SOMs have already been used in Computational Phonogram Archiving (Blaß & Bader, 2019), e.g., defining ethnic groups (Bader et al., 2021).

By training the SOM with all musical pieces and placing the pieces at the best-match positions on the SOM map, clusters might be found concerning musical content. The SOM is used instead of other ML methods, like deep-learning or convolutional neural networks, as the task is analysis. While most ML methods perform well, after successful training, the trained weights do not trivially display the reason for appearing clusters. Then, no analytical results can be derived. On the contrary, a trained SOM consists of a set of component planes, which are 2D maps for each feature of the trained vector. These component planes deliver the analysis results. Furthermore, the distance or u-matrix between neighboring neuron distances, calculated as Euclidian distances, displays regions of similarly trained neurons and, therefore, helps detect clusters.

---

[17] The frequencies and intervals in cent of Kyai Kanjut Mesem were measured by Surjodiningrat et al., (1972) in their study on the tone measurements of outstanding Javanese gamelans.

The input to the SOMs in the present study is the results of four timbre analysis algorithms calculated from the recording sound. To arrive at these four algorithms, in a pre-study, many other psychoacoustic features have been used, like roughness, fractal correlation dimension as a measure of music complexity, or loudness, all calculated over time frames of $2^{14}$ sample points and summarized as mean or standard deviations over all frames of the single pieces. Many different combinations have been used as input feature vectors to the SOM, and the results have been judged in terms of the appearance of clusters. The psychoacoustic features described below were found to arrive at such clusters best. This does not mean they must be the only way to cluster the pieces; no reasonable clusters were found using other features.

The main topic of this investigation is whether Indonesian and Western ensembles are considerably different. They might differ in instrument sounds, tonal systems, articulation, or large-scale form. Of course, they might also differ in ensemble setup, musical repertoire, or other features, which are beyond the scope of this investigation. To estimate the three parameters, different analyses were performed:

1) **Tuning.** In Section 5, it was discussed how tuning gamelan instruments affects pitch and timbre simultaneously. Researchers agree that this relation is important for musical perception, even though they disagree on how the perception is affected (Marjieh et al., 2024; Wendt & Bader, 2019). In this study, the spectrum was calculated by FFT with a frequency range of six octaves from 20 Hz to 1280 Hz. As gamelan ensembles most often have decisively different tonal systems, such systems are expected to appear in the spectrum, along with the different timbre of particular instruments.
2) **Articulation:** The most prominent perceptual timbre feature is the spectral centroid. Therefore, temporal changes in spectral centroid are a compositional method of articulation (Bader, 2013). In the study, the standard deviation of the spectral centroid is used to detect articulation. The same reasoning holds for the spectral spread, displaying how dense a timbre is and how many instruments play simultaneously. Therefore, the standard deviation of the spectral spread was also used. Third, sharpness can also be used for articulation. Here, the mean sharpness calculated over single pieces was used.
3) **Musical large-scale form:** For describing composition on a gong cycle or several gong cycles, the same algorithms for articulation were used. The spectra of the temporal developments of these timbre parameters were calculated to tell articulation from large-scale form and the mean and standard deviation. In these spectra, articulation is expected to show peaks around 0.5 Hz, while large-scale form shows up at much lower frequencies. A spectral centroid was calculated to arrive at a single number for these spectra.

The COMSAR and APOLLON framework were used for this purpose[18].

---

[18] https://github.com/ifsm

# 8 Results
## 8.1 Tuning

The SOM trained for gamelan tuning clusters the gamelan ensembles with respect to the musical pieces used in the study. Fig. 1 shows the trained SOM with the pieces used for training plotted at the best-matching positions on the map. The background color is the u-matrix, displaying similarities or dissimilarities between neighboring neurons, where dark colors are regions of high similarity and light colors are those of high dissimilarity. In the figure, each gamelan ensemble has a unique color, and each piece played by one ensemble is represented by a circle with the color associated with this ensemble. The Indonesian ensembles have round circles, while the Western ensembles have diamonds.

Clearly, each ensemble is clustered at a unique region on the map. Yellow ridges on the u-matrix separate these regions. So, e.g., the Indonesian Gamelan Mankuagaran clusters at the left lower side and is surrounded by a yellow ridge while lying in a dark blue region. Therefore, these pieces are very similar with respect to their spectra, so their tonal systems differ considerably from other ensembles.

Still, one of the pieces of this ensemble is located at the upper right corner between the Bali New Zealand (Taniwha Jaya) and the Margi Bodoyo ensemble, disturbing the cluster. Also, the Arum Sih ensemble pieces are split into two groups, one in the lower left and one in the lower right corner. These pieces differ considerably from other ensembles but also within one ensemble. Such exceptions are of interest but still beyond the scope of this paper. The clusters are clearly present, and their tonal systems can detect the ensembles.

On the other hand, no clusters for Indonesian and Western ensembles show up. Therefore, we conclude that Indonesian and Western ensembles do not systematically differ in tonal systems.

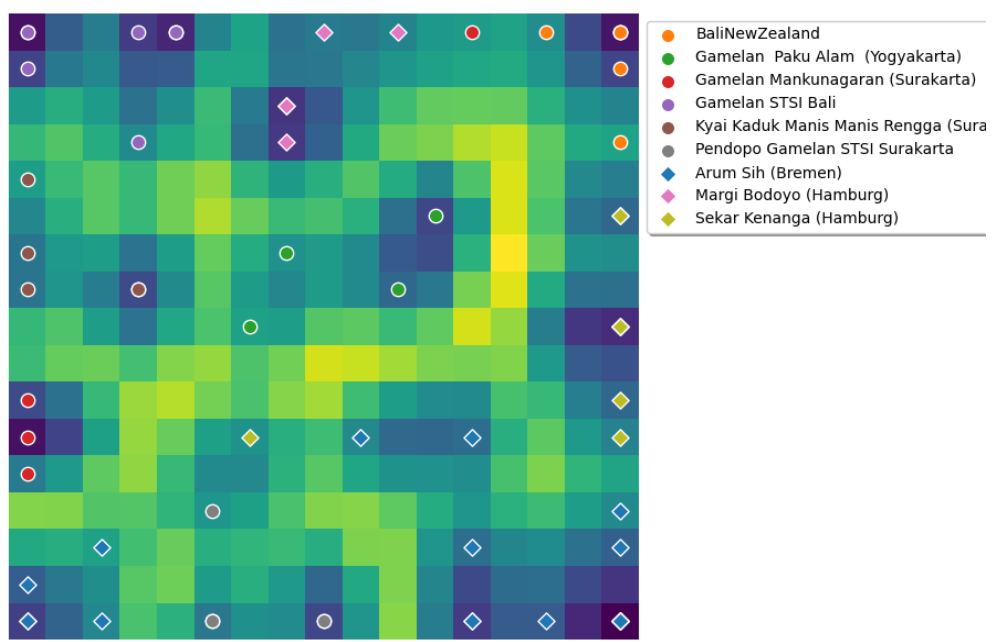

**Fig. 1:** SOM mapping of gamelan pieces of different Indonesian and Western ensembles trained by FFTs of the pieces and best fit on the trained map. The ensembles cluster in terms of their unique tonal systems and timbre. Still, no clusters between Indonesian and Western ensembles appear.

## 8.2 Articulation / Large-scale form

The trained SOM in Fig. 2 shows the results of the timbre features spectral centroid (standard derivation), spectral spread (standard derivation), and sharpness (mean value) and the best-fit positions of all pieces the map was trained with. These features can detect both articulation and large-scale form.

Clearly, the Indonesian pieces (circles) are placed on the upper left half of the plot, while the Western pieces (diamonds) are on the lower right. The u-matrix in the background also shows stronger differences between the pieces on the upper left side. Therefore, the Indonesian pieces differ stronger in articulation or large-scale compared to the Western pieces. At first, this does not mean that the pieces themselves have strong articulation but only that pieces differ within this parameter.

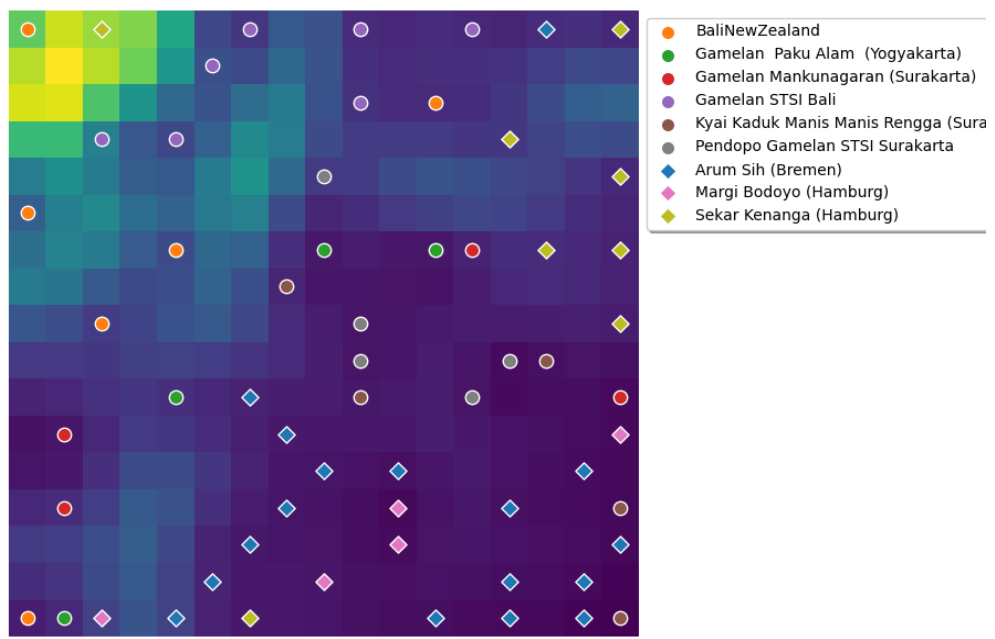

**Fig. 2:** SOM and best-fit positions of gamelan pieces using the articulation and large-scale form timbre features as input vector. Indonesian and Western pieces cluster, while Indonesian pieces are more located on the upper left and Western ones on the lower right side. The u-matrix shows Indonesian pieces are more differentiated compared to Western pieces.

To arrive at the reason for such clustering, in Fig. 3, the component planes of the three features are shown. The standard deviation of the spectral centroid and the spectral spread both show large values in the upper left corner and relax towards the lower right corner. The spectral spread also has larger values at the upper right corner. The mean of the sharpness again shows strong values at the upper left corner while stronger values at the lower left corner also appear.

This analysis confirms that the Indonesian pieces show much stronger articulation and/or large-scale form differentiations than Western pieces.

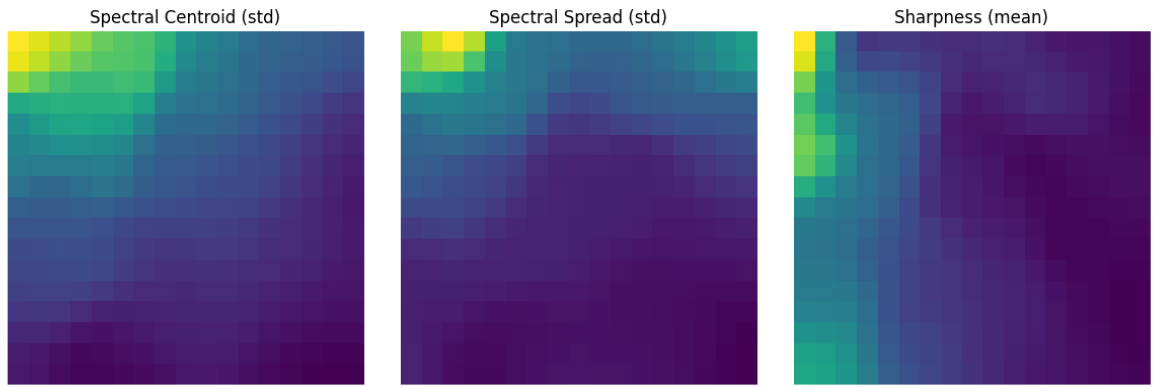

**Fig. 3** Component planes of the SOM of Fig. 2 for articulation / large-scale form showing Indonesian pieces having more articulation and/or large-scale form variations than Western pieces.

To estimate if the clustering between Indonesian and Western gamelan ensembles is caused by articulation or large-scale form, in Fig. 4, the most prominent peak frequencies in the spectrum of the spectral centroid timbre time development are plotted for all pieces sorted by ensembles. If the spectrum of the spectral centroid timbre feature has large peaks at very low frequencies, large-scale form is strongly present, and therefore, large-scale form will be at least one cause for the clustering.

Indeed, most ensembles have the most prominent peak at very low frequencies with a median of 0.01 Hz, corresponding to 100 s time intervals. The only exception is the Arum Sih ensemble. Next to values as low as the other ensembles, it also shows the most prominent peaks for two pieces, *Udan Mas* and *Gendrowo Momeng*. Aural inspection indeed shows no clear large-scale compositional elements for these pieces. Therefore, this method seems also to be a detector for the presence of a large-scale form.

Fig. 5 shows the spectral centroid of the spectra of the spectral centroid timbre feature. These values are considerably higher than the most prominent peaks of Fig. 4, pointing to articulation. Again, the Arum Sih ensemble has the highest values, although not the largest spread of these values. There is a tendency for the Western ensembles on the left side of the plot to have higher values above the median of 0.52 compared to the Indonesian, with lower ones more below the median.

Note that the Arum Sih pieces as Western ensembles show low values of centroid standard deviation in the SOM component planes of spectral centroid and spread standard deviations and are located at positions with low u-matrix values, showing a high consistency between pieces; they show the largest spread in the plot for most prominent peaks. This is possible as both are independent measures. For those pieces, changes in the timbre of the spectral centroid over time are small. This can happen with pieces of a clear large-scale form (frequencies around 0.01 Hz of most prominent frequency) or with a free form (frequencies around 0.7 – 0.8 Hz).

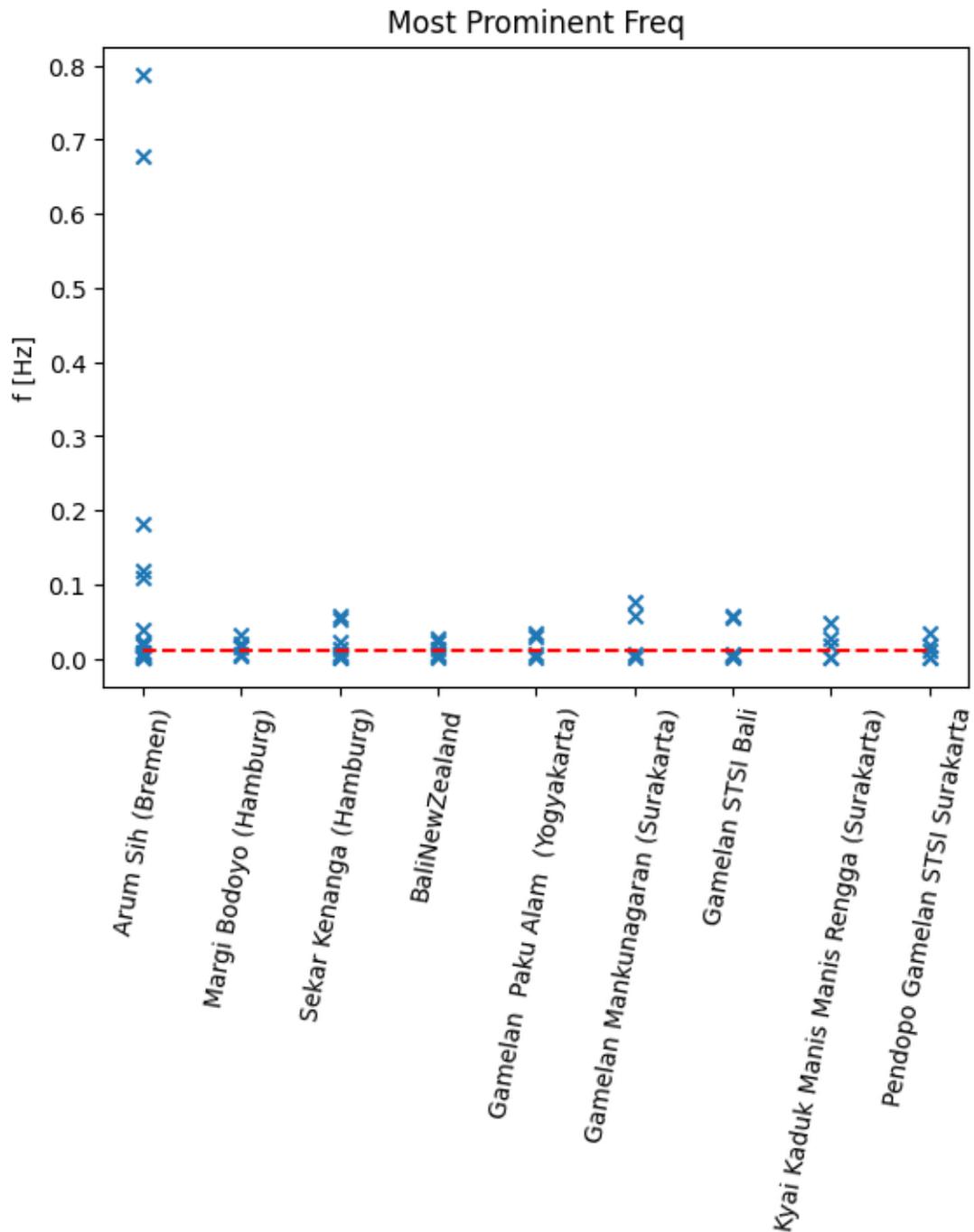

**Fig. 4:** Most prominent frequencies of the spectra of the temporal development of the spectral centroid timbre parameter for all pieces. Most ensembles are around 0.01 Hz – 0.04 Hz, corresponding to time intervals of 100 s – 25 s, therefore pointing to a large-scale form. The red line is the median of 0.01.

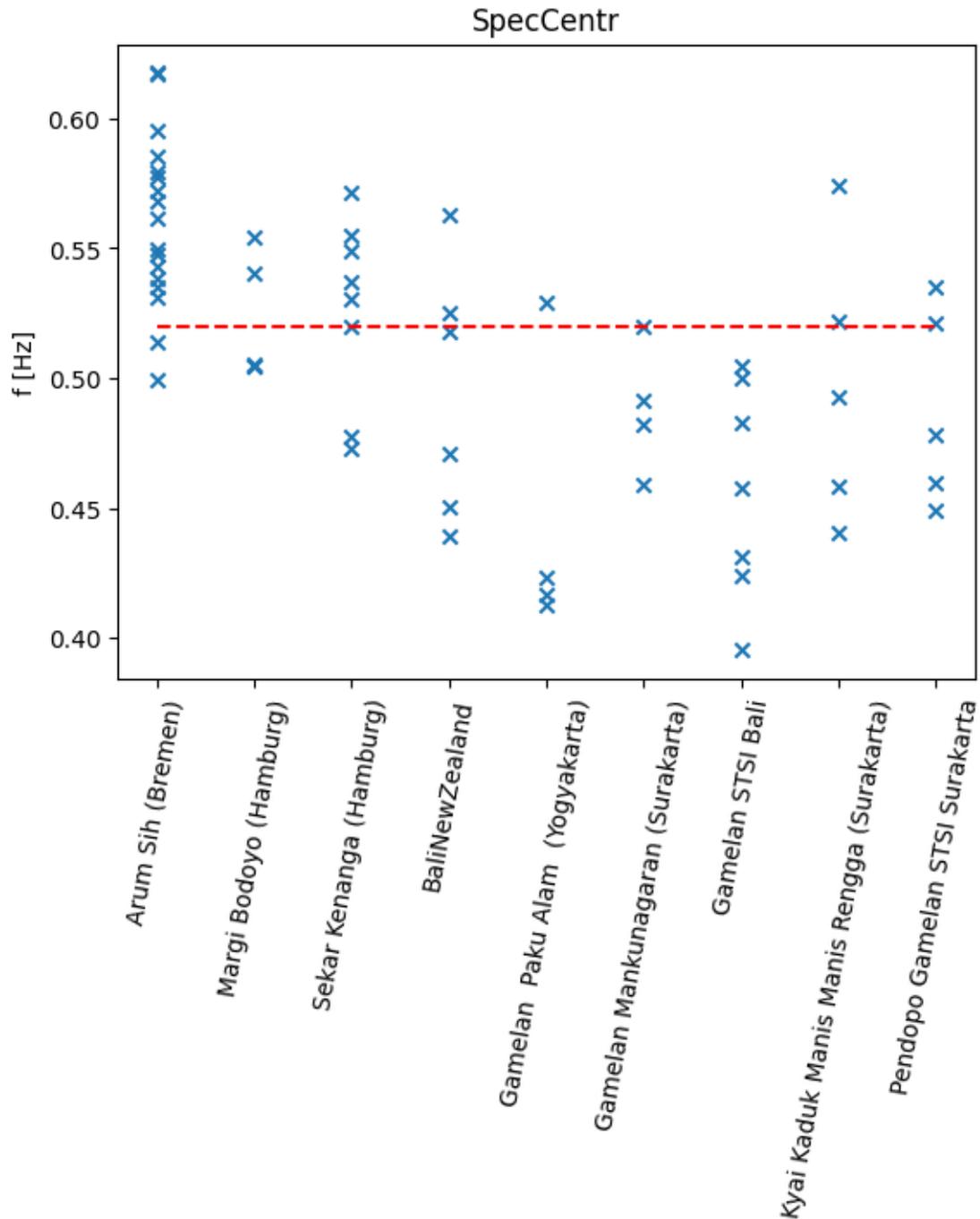

**Fig. 5:** Spectral centroid of the spectrum of the spectral centroid timbre feature showing the main distribution of peaks. These values are much larger, around 0.5 Hz, corresponding to 2 s intervals, compared to the most prominent peaks, pointing to articulation. The red line is the median of 0.52.

# 9  Conclusions

At least concerning the parameters used in the present investigation, Indonesian gamelan ensembles differ considerably from Western gamelan ensembles with regard to performance parameters articulation and large-scale form. Thereby, Indonesian ensembles have stronger articulation and a larger variation in their large-scale form.

Taking into consideration the transfer of gamelan into the West and the reception of these ensembles as exotistic, where the West often imagines a soft and meditating state of musical performance practice and perception, a reduction of articulation and a more compact large-scale form corresponds to such idles. The finding of more improvised forms in the Arum Sih ensemble, which lack a clear large-scale form, also points in this direction. Another aspect might be the use of gamelan. While in Indonesia, the ensemble is often used to accompany a shadow play with decisive elements of dramatic actions, in the West, gamelan is most often performed in a concert form lacking such dramatic elements. Furthermore, the choice of musical pieces and their interpretation might explain those differences. Gamelan ensembles in the courts, as they are featured in this study, often play musical suites with multiple changes over time in musical density and dynamics. On the other hand, non-Indonesian gamelan ensembles often play much shorter pieces characterized by more constant musical features due to a lack of experienced players or the different performance habits in the West.

## 5.1 Discography